# Pathological myopia classification with simultaneous lesion segmentation using deep learning

Authors:

Ruben Hemelings[a,e]*, MS

Bart Elen[e], MS

Matthew B. Blaschko[c], PhD professor

Julie Jacob[b], MD PhD

Ingeborg Stalmans[a,b], MD PhD professor

Patrick De Boever[d,e], PhD professor

Affiliations:

[a] Research Group Ophthalmology, KU Leuven, Herestraat 49, 3000 Leuven, Belgium

[b] Ophthalmology Department, UZ Leuven, Herestraat 49, 3000 Leuven, Belgium

[c] ESAT-PSI, KU Leuven, Kasteelpark Arenberg 10, 3001 Leuven, Belgium

[d] Hasselt University, Agoralaan building D, 3590 Diepenbeek, Belgium

[e] VITO NV, Boeretang 200, 2400 Mol, Belgium

*corresponding author

Contact details

Affiliation: KU Leuven, VITO

Postal address: Vito Health, Industriezone Vlasmeer 7, 2400 Mol, Belgium

E-mail: ruben.hemelings@kuleuven.be

Phone: +32472748707

Abstract

Purpose: To assess the use of deep learning for detection of pathological myopia (PM) and semantic segmentation of myopia-related lesions from fundus images.

Methods: This investigation reports on the results of deep learning models developed for the recently introduced Pathological Myopia (PALM) dataset, which consists of 1200 images. Evaluation metrics include area under the receiver operating characteristic curve (AUC) for PM detection, Euclidean distance for fovea localization, and Dice and F1 metrics for the semantic segmentation tasks (optic disc, retinal atrophy and retinal detachment). We also introduce a new Optic Nerve Head (ONH)-based prediction enhancement for the segmentation of atrophy and fovea.

Results: Models trained with 400 available training images achieved an AUC of 0.9867 for PM detection, and a Euclidean distance of 58.27 pixels on the fovea localization task, evaluated on a test set of 400 images. Dice and F1 metrics for semantic segmentation of lesions scored 0.9303 and 0.9869 on optic disc, 0.8001 and 0.9135 on retinal atrophy, and 0.8073 and 0.7059 on retinal detachment, respectively.

Conclusions: We report a successful approach for a simultaneous classification of pathological myopia and segmentation of associated lesions. Our work was acknowledged with an award in the context of the "Pathological Myopia detection from retinal images" challenge held during the IEEE International Symposium on Biomedical Imaging (April 2019). Considering that (pathological) myopia cases are often identified as false positives and negatives in glaucoma deep learning models, we envision that the current work could aid in future research to discriminate between glaucomatous and highly-myopic eyes, complemented by the localization and segmentation of landmarks such as fovea, optic disc and atrophy.

Introduction

Myopia or nearsightedness currently affects approximately 34% of the world population.[1] High myopia, often defined as a spherical equivalent that exceeds -6.00 diopter or an axial length of 26.5mm or more, has a prevalence ranging from 1% in African Americans[2] and up to 5.5% in the Japanese Tajimi study[3]. Approximately

1-3% of the world population develops vision-impairing macular lesions (lacquer cracks, choroidal neovascularization, and Fuchs spots) as a result of high myopia, referred to as myopic maculopathy.[4,5] Both the presence of myopic maculopathy and posterior staphyloma are used to define pathological myopia (PM), which causes uncorrected and irreversible visual impairment.[6] Other retinal changes due to myopia include: fundus tessellation, (peripapillary) atrophy, optic disc tilting, retinal tear and retinal detachment. Additionally, myopia increases the risk of developing open-angle glaucoma[7], presumably because myopic eyes have thinner and weaker lamina cribrosa tissue[8]. Optic nerve head (ONH) changes such as temporal disc flattening and tilting[9], as a consequence of myopia, hampers glaucoma detection through ONH assessment during fundoscopy or fundus image analysis[10]. Peripapillary atrophy (PPA), being attenuation of retinal pigment epithelium (RPE) neighboring the ONH, is associated with both myopia and glaucoma, and is one of the causes for a high number of myopic patients being diagnosed as glaucoma suspects.

Artificial intelligence (AI), especially deep learning (DL) is showing great potential in ophthalmic research for disease identification and staging[11]. Relevant for this manuscript is refraction estimation from fundus images using deep learning by Varadarajan et al. (2018), who developed a regression model that estimates refractive error with high accuracy (<1 diopter mean absolute error).[12] Semantic segmentation of fundus images has been applied for vessel extraction[13], artery/vein discrimination[14], and optic cup/disc estimation[15]. Segmentation of myopia-related lesions from fundus images and deep-learning based classification of pathological myopia has not been previously explored.

Here, we report the methods and results developed for the classification of (non-)pathological myopia, fovea localization, and semantic segmentation of optic disc, retinal atrophy and detachment.

Methodology

Dataset and evaluation

Retinal images were made available in the context of the "Pathological Myopia detection from retinal images" challenge held on the occasion of the IEEE International Symposium on Biomedical Imaging organized in April 2019.[16] The PALM dataset consists of 1200 anonymized color fundus images that were captured with a Zeiss VISUCAM device at a 45° angle with a 2124 x 2156 resolution or 30° with a 1444 x 1444. The images are

macula- or optic disc-centered of left eyes with no disclosure of the number of different eyes or patients that were included in the dataset. The 1200 images are split into equally sized train, validation, and test sets sharing the same characteristics. Publicly available labels for the training set of 400 images encompass (1) the binary label for (non-)pathological myopia classification, (2) cartesian coordinates corresponding to the location of the fovea, and (3) semantic segmentation ground truth on pixel level for optic disc, peripapillary/retinal atrophy and retinal detachment. The myopia labels were extracted from the health records of the Zhongshan Ophthalmic Center, Sun Yat-sen University (China) and were determined during an ophthalmic examination, including optical coherence tomography (OCT) and visual field (VF) testing. The fovea coordinates and segmentation masks were generated by seven independent ophthalmologists from the same clinic. The PM detection training labels are balanced (53% PM images), but do not match the prevalence encountered in screening context (up to 3%). Ground truth of optic discs is available for most images, with an empty ground truth mask in case of an absent or partially visible disc. An overview of official training set characteristics is provided in Table 1. Differences in PM and non-PM characteristics were analyzed using a two-tailed t-test.

PM detection was quantified using area under the receiver operating characteristic (AUC), while the fovea localization was evaluated using the average Euclidean distance between the predicted cartesian coordinates and ground truth. The three predicted segmentation masks (optic disc, atrophy, detachment) were evaluated using a weighted combination of Dice[17] similarity coefficient (segmentation) and test's accuracy using the F1 score (detection). See supplementary information for full details on evaluation framework as defined by PALM organizers.

Network architectures and loss functions

UNet++[18], a nested variant of the widely used U-Net[19], was selected for the segmentation tasks because of its reported improved performance. The widely used ResNet[20] encoders were tested as feature extractors to enable transfer learning with pretrained ImageNet[21] weights. At the end of the contracting path, where the input image is converted to a representation in latent space, we added a second output branch for PM classification in light of co-regularization.[22] Figure 1 displays the typical hourglass architecture, with the contracting path extracting and refining feature maps through convolutional, batch normalization and pooling layers. In UNet++, these feature maps are connected to a number of dense convolution blocks, before being inserted in the expanding path (decoder).

The employed loss function for PM classification is standard binary cross-entropy. Fovea localization labels are cartesian coordinates, but were converted to filled circles with varying radii to allow for segmentation, as an alternative approach to standard coordinate regression. All segmentation models employed standard binary/categorical cross-entropy as loss function, complemented by Dice similarity coefficient. Finally, we experimented with the Lovàsz-Softmax[23] as third loss component. The latter serves as a tractable surrogate for the optimization of intersection over union (IoU), and has proven itself as finetuning loss in recent semantic segmentation challenges.[24]

Preprocessing, Data augmentation, Training details

Color fundus images are unevenly illuminated due to the curvature of the retina. Local contrast enhancement through background subtraction estimated by a large Gaussian kernel was used to correct this[25]. Data augmentation techniques used throughout all experiments include random cropping, mild elastic deformation, and horizontal flips. Random cropping was performed selecting patches of 288 x 288 within resized images of a random size between half and original image size to teach the model features at multiple resolutions. The model input of 288 x 288 was selected based on a balance between the merits of pretrained weights (224 x 224) and segmentation output (higher resolution leads to better results).

Due to the severe class imbalance of the retinal detachment segmentation, we adopted a sampling strategy that oversamples images with retinal detachment at earlier stages of the training process to an equal mini-batch distribution, only to gradually slim down to the original data distribution (x 0.75 per five epochs). As such, the model is less likely to treat the detachment label as noise at training start.

Model development was done in Keras v2.2.4 with TensorFlow v1.4.1 backend. All models used Adam[26] optimizer with a default starting learning rate at 0.001. A plateau callback decreased the learning rate by 25% after ten successive epochs of stagnation in validation metric (Dice). To obtain a wider optimum, model weights were averaged over the last twenty epochs when the learning rate reached a value of $1e^{-5}$.[27] Internal validation was performed on a holdout set of 40 images, representing 10% of available training data.

ONH-based prediction enhancement

Theoretically, there should be no overlap between atrophy and optic nerve head (ONH). Peripapillary atrophy represents loss of RPE and choriocapillaris, which ends/starts in Bruch's membrane opening (BMO), and simultaneously delineates the optic disc boundary. Leveraging this domain knowledge, the optic disc and peripapillary/retinal atrophy segmentation tasks were bundled by fusing the two ground masks. Retinal detachment ground truth does overlap with atrophy in certain cases, hence this ground truth was left unprocessed.

In addition to standard coordinate regression, we rebranded the fovea localization task as a segmentation problem. The ground truth masks were generated by drawing filled circles (varying radii) based on the official cartesian coordinates as centroids. The optic disc is located on the nasal side of the fovea. Hence, the optic disc segmentation ground truth was added to the fovea ground truth, to implicitly insert this domain knowledge. We also experimented with the implementation of cutout[28], a common regularization technique, to improve the learning of the ONH – fovea relation.

The predicted fovea segmentations required post-processing in case of missing or unlikely predictions. Two sanity checks were performed prior to reconversion to coordinates: (1) whether there is a fovea prediction made, and (2) whether it falls within normal range compared to optic disc location. Normal range was defined as mean ± 2 x standard deviation, with population mean and deviation estimated from the training labels (grouped by image resolution). If the assertions failed, the predicted fovea coordinates were determined based on optic disc centroid and mean distance between optic disc and fovea. For benchmarking purposes, we also report on experiments without joint optic disc segmentation. Here, the postprocessing was limited to the use of image center coordinates in case of missing fovea prediction.

Ensembling on image and model level

Ensembling on image and model level tend to lead to small performance gains due to its decrease in prediction variance. Hence, final predictions of (non-)pathological myopia classification were obtained through commonly-used test-time augmentation techniques (elastic deformation and horizontal flips), complemented by ensembling on the model level (7 ResNet models with different random seed on train/val split). Segmentation results were generated using averaged predictions on overlapping 288 x 288 patches from resized images (288 x 288, 294 x 294, and 302 x 302). Overlapping patches were only possible in the last two resolutions.

Results

Table 1 reveals that the largest group of available training images are 45° macula-centered images, whereas its disc-centered variant contains only 3 images. Complete optic discs are missing in all 30° macula-centered images, and in some PM cases imaged at 45° as well. Optic disc area ranged between 1-4%, and was significantly larger in 30° disc-centered PM images. Retinal atrophy was present in almost all PM cases, and in roughly half of non-PM images. The area covered by atrophy was larger in PM images for all modalities. The fovea is visible in nearly all images.

Table 2 provides an overview of ablation experiments on a holdout set of 40 images of the training set. All PM cases were correctly classified in both experiments, but the validation loss was significantly lower in the setup with combined ONH and atrophy segmentation (0.0824 versus 0.1146). The Dice score on ONH segmentation was found to be the highest in the vanilla setup with a single model (0.9481 Dice). For retinal atrophy however, multi-class segmentation with Lovász as loss component did lead to better performance (0.6948 Dice) when compared to two individual models (0.6210 Dice). The balanced data generator did lead to better performance in segmentation of retinal detachment (0.9998 Dice). The move from regression to segmentation for fovea localization seems to be beneficial, with average Euclidean distance at 229 and 129 pixels, respectively. The result using a segmentation approach also improved when employing a larger fovea radius (110 pixels Euclidean distance). Our proprietary ONH-based prediction enhancement led to a major performance gain (87 pixels). Finally, the post-processing that deals with missing and unrealistic predictions resulted in the best observed performance (62 pixels).

Table 3 summarizes the quantitative results on the official validation and test set, obtained through the online competition evaluation server hosted at http://ai.baidu.com/broad/subordinate?dataset=pm. The trained models for detection of pathological myopia achieve AUC values at 0.9934 and 0.9867 on validation and test set, respectively. The results on fovea localization differ significantly between the two holdout image sets (79.42 versus 58.27 Euclidean distance), which is caused by the implementation of the ONH-based prediction enhancement between validation and test prediction in the PALM competition. The mean gap of approximately 20 pixels indicates that the post-processing does improve significantly on the vanilla output of the deep learning models (in line with ablation study, Table 2). Dice and F1 metrics for both optic disc and atrophy segmentation are consistent between validation and test set. Finally, the F1 metric for retinal detachment segmentation reveals that the validation and test set contain 12 and 11 cases of retinal detachment, respectively. The trained deep learning models identified six correct cases in both sets.

Ground truth for validation and test sets on image level will be made publicly available at a later date by the organizers of the PALM challenge. Hence, the qualitative results of four test images displayed in Figure 2 cannot be visually compared to the official ground truth. The optic disc – outlined in green – is detected in both non-pathological (A) and pathological (B,C) fundus images (not present in D), and does not overlap with peripapillary atrophy (B,C). The fovea – indicated by a cross – is localized well in cases of a clear (A) and covered (C,D) macula, or added during postprocessing (B). Atrophy – outlined in white – is segmented at both peripapillary (A,B,C,D) and macular (B) regions. In images where 30% of the image is predicted to be retinal detachment, the prediction is replaced with the size of the image mask (yellow outline of image C).

Discussion

This deep learning study on fundus images describes (1) the detection of pathological myopia (PM), (2) the localization of the fovea, and (3) the segmentation of optic disc, retinal atrophy and retinal detachment. The results are obtained after training on 400 labeled fundus images and relies on transfer learning and co-regularization through weight sharing. The methodology described in the manuscript led to a third place in PALM challenge hosted at ISBI 2019. The PALM dataset provides novel challenges to existing research topics, as myopic optic discs are often tilted (optic disc segmentation), and the fovea obscured due to tessellation and macular atrophy in some cases of pathological myopia (fovea localization).

The PM detection task scored an AUC of 0.9867 on the official test set of 400 images. PM detection from fundus images has not been covered in deep learning literature prior to the launch of PALM. The work of Varadajaran et al (2018) comes closest, but employs a whole different setup. Their goal was to develop a data-driven regression model that estimates refractive error (including cases of pathological myopia), using the spherical equivalent as target. One important insight from their work is the saliency concentration surrounding the macular area. Some possible explanations are given, such as the variance in foveal focus and reflectance in fundus images, but remain unexplored hypotheses.

In our investigation, the task of PM detection was approached in a different manner, given the different nature of the task and materials. The definition of PM states that a highly myopic case is converting to pathological once a posterior myopia-specific pathology from axial elongation is developing, such as vision-impairing myopia-induced lesions. This is corroborated by the explorative analysis of the training set, given in Table 1. Retinal atrophy, being progressive RPE thinning and attenuation, is present in 98.3% cases of PM, versus 52.6% in non-PM images (restricted to the modality of 45° macula-centered images). By combining atrophy segmentation and PM classification, one forces the AI model to focus on lesions as main features that contribute to PM classification. This implies a step towards explainable AI or sufficient transparency to gain clinicians' trust in the future use of deep learning detection systems in ophthalmology.

The optic disc segmentation model obtained a Dice similarity coefficient of 0.9303, scoring in line with relevant work.[29] Due to axial elongation, myopia induces anatomical changes to the optic nerve head, resulting in tilted and oval-shaped optic discs, often surrounded by peripapillary atrophy. These alterations are significant, as a pretrained optic disc segmentation model on non-myopic fundus images failed to properly delineate the discs in the PALM dataset. Another factor could be the larger optic disc size observed in myopic eyes[30,31]. From Table 1, there is a moderate significance ($P < 0.01$) found between PM and non-PM (which also includes high myopia) in the 30° disc-centered images. Hence, optic disc size is unlikely to be an informative predictor in PM detection.

This original investigation also introduces a pioneering result of 0.8001 Dice on the segmentation of retinal atrophy (PPA, lacquer cracks and Fuch's spots) in fundus images. This type of segmentation can support future research in discriminating between myopia- and glaucoma-induced peripapillary atrophic changes. This is relevant because in previous work it has been observed that false positive and negative predictions in glaucoma classification models are often due to cases of high/degenerative myopia. For example, Liu et al (2019) observed that the most common reason for both false-negative and false-positive grading by their DL model (46.3% and 32.3%) and manual grading (44.2% and 34.0%) was pathological or high myopia.[10] Several studies investigated the discriminatory properties of beta- (area with intact Bruch's membrane) and gamma-PPA (lacking Bruch's membrane) for myopia and glaucoma using OCT, but report contradictory findings and low discriminatory power.[32,33] Another recent study discovered a relationship between PPA shape and glaucoma progression, stating that progression is more correlated with eccentric PPA than concentric PPA.[34] DL may assist in analyzing PPA in a larger set of patients than previous investigations.

The fusion of optic disc and atrophy segmentation tasks ensured no overlap in final predictions. This form of joint prediction increases the odds of generalization to unseen samples (in this case, 800 images split in validation and test set of equal size). Ground truth fusion did lead to better performance for atrophy segmentation, but not for ONH segmentation. Another important motivation for joint training is explainable artificial intelligence, as previously discussed.

For fovea localization, we initially considered adding a regression branch to the segmentation model for optic disc and retinal atrophy. However, due to subpar performance (229 pixels Euclidean distance), this idea was discarded and replaced by a standalone segmentation model. One potential explanation for poor regression performance could be the combination of scarcity in available regression labels (1 per image) when compared to segmentation labels (1 per pixel), and low variance in coordinate values (the fovea is centrally located in macula-centered images). The main disadvantage of a segmentation approach is the loss of direct optimization on the competition metric. The performance gain in Euclidean distance between fovea localization results on official validation (79.42, without post-processing) and test (58.27) demonstrates the importance of thoughtful post-processing. Fovea localization in fundus images has been investigated with deep learning prior to PALM, but primarily in clean datasets with clear macular depression.[35] Our domain knowledge insertion – combined optic disc and fovea segmentation – is considered useful in the move towards general deep learning models that can process large amounts of fundus images with unclear macular regions. Such fovea localizing models can assist future big data research. One application would be the automated image cropping of the macula area to facilitate diabetic retinopathy screening.

Finally, this study reports a Dice score of 0.8173 on the task of retinal detachment segmentation. The high performance is mainly due to the correct predictions of empty masks in the high number of cases (~97% of images) without retinal detachment. The actual performance would be much lower when the images with retinal detachment would be isolated. In most cases, retinal detachment covers more than half of the field of view (FOV) in the fundus. Hence, one could question the added value of segmentation over a classification approach.

This study suffers from several limitations. Given that the images, made with a single device, were sourced from a Chinese hospital, the ethnicity is likely to be predominantly East-Asian. Further evaluation on other ethnicities and fundus camera devices is needed to ensure model generalizability. Additionally, the ground truth on image level for validation and test sets are currently unavailable, hampering the qualitative comparison of semantic segmentation results, and the calculation of specificity and sensitivity. On the other hand, the introduction of medical labeled datasets and robust online evaluation server should be encouraged, as they allow the objective comparison of innovations in deep learning for medical imaging.

Conclusions

We report a successful approach for a simultaneous classification of pathological myopia and segmentation of associated lesions. These award-winning results were obtained in the context of the "Pathological Myopia detection from retinal images" challenge held on the occasion of the IEEE International Symposium on Biomedical Imaging organized in April 2019. Considering that (pathological) myopia cases are often found as false positives in glaucoma deep learning models, we envision that the current work could aid in future research to discriminate between glaucomatous and highly-myopic eyes, complemented by the localization and segmentation of landmarks such as fovea, optic disc and atrophy.

Acknowledgements

The first author is jointly supported by the Research Group Ophthalmology, KU Leuven and VITO NV. This research received funding from the Flemish Government under the "Onderzoeksprogramma Artificiële

Figures

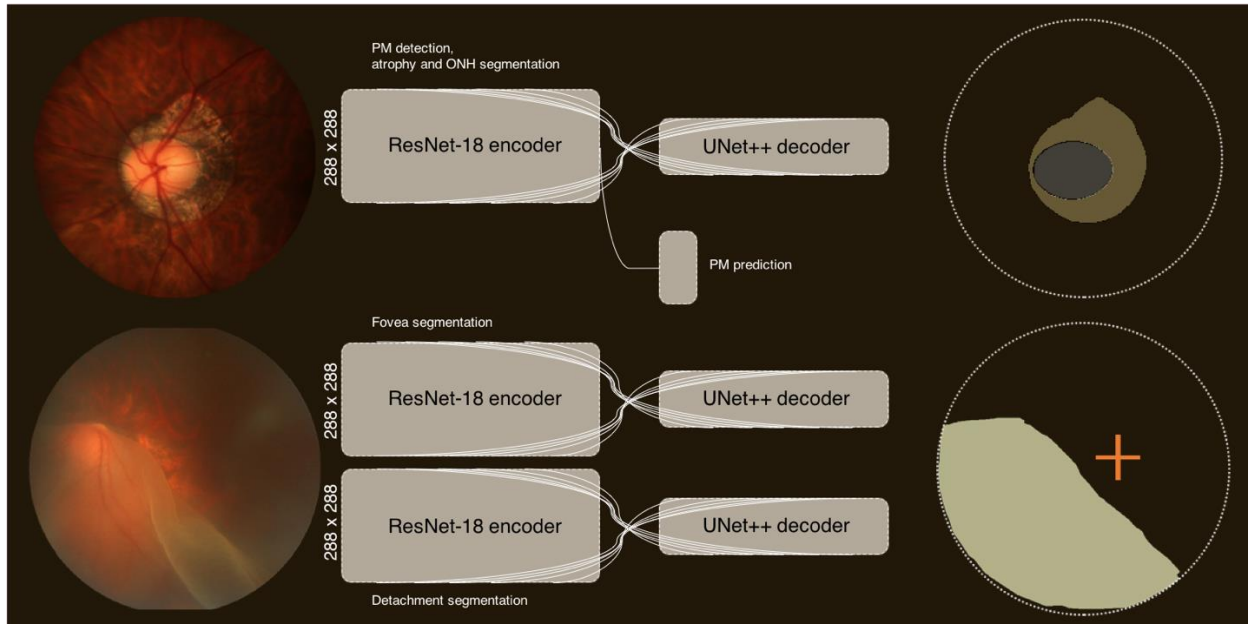

*Figure 1: Overview of the final three models used for inference on the PALM official validation and test set. The first model is aimed at PM classification with simultaneous segmentation of ONH and retinal atrophy. The ResNet encoder accepts resized fundus images of (288 x 288) and outputs (9 x 9 x 512) at the latent space. The decoder upscales this output back to the original image size, using a plethora of skip connections (illustrated by white lines). The graphic on the upper right represents the generated segmentation map of the ONH (grey) and retinal atrophy (olive). The output of the encoder is also separately transformed to a single prediction for PM classification (see supplementary material for layer details). The model for fovea localization employs a similar architecture as for ONH/atrophy segmentation, but generates a circle. This circle is then transformed to coordinates using its centroid (visualized by the orange cross on the right bottom segmentation map). Finally, the UNet++ model for segmentation of retinal detachment is identical to the other models, but outputs detachment.*

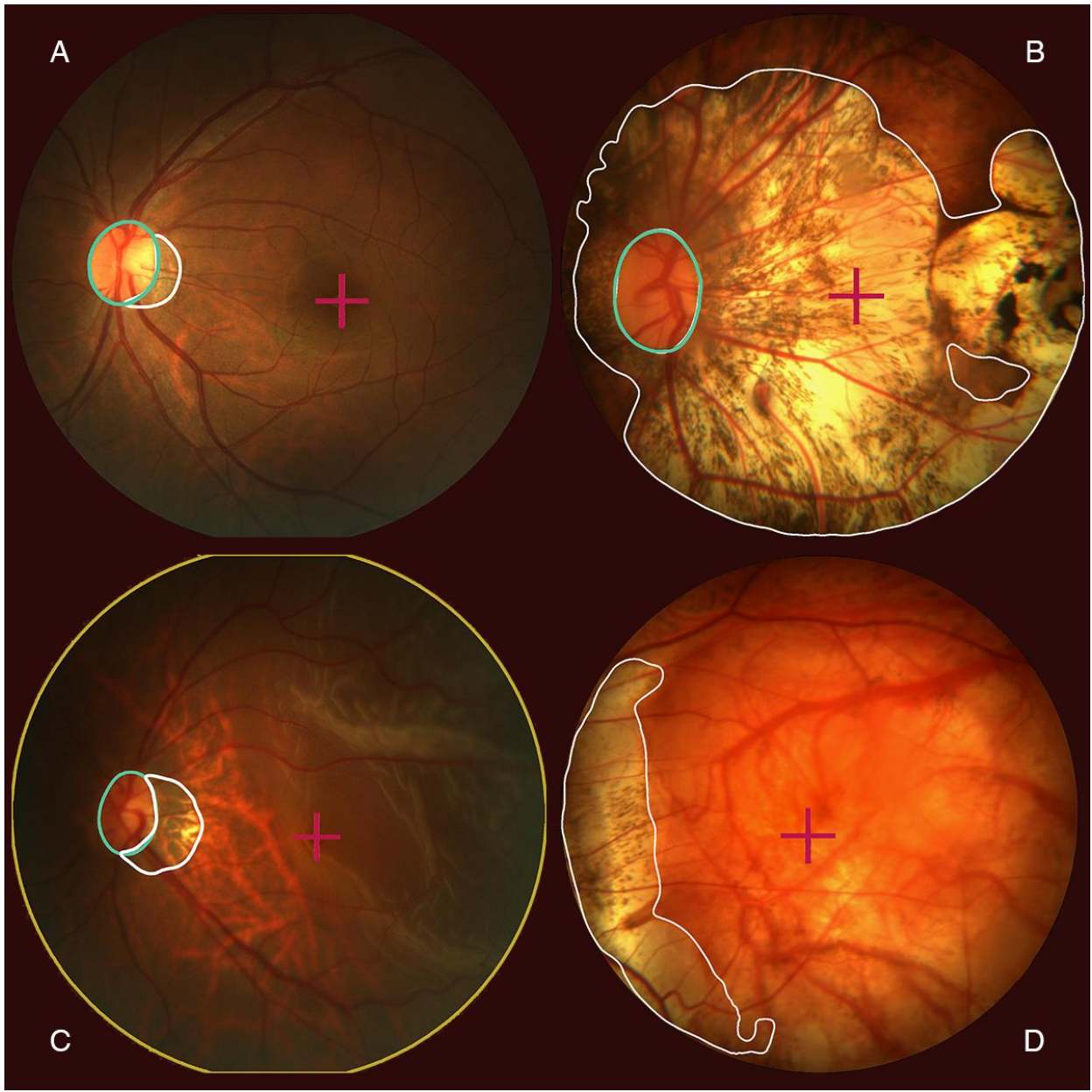

*Figure 2: Qualitative results giving four cases of the official test set. The optic nerve head (outlined in green) is detected and segmented in A, B and C. Retinal atrophy is detected and segmented (outlined in white) in B and C. Retinal detachment was detected in C, for which the whole fundus is outlined in yellow. Finally, the fovea is localized in all cases, indicated by a purple cross.*

Tables

*Table 1: Overview of characteristics of labeled training set of 400 images. Significance level between PM and Non-PM on same camera settings provided with asterisks (where applicable, * <0.05, ** <0.01, *** <0.001, **** <0.0001).*

| Centering | Macula | | | | Disc | | | |
|---|---|---|---|---|---|---|---|---|
| Angle | 30° | | 45° | | 30° | | 45° | |
|  | PM | Non-PM | PM | Non-Pm | PM | Non-PM | PM | Non-PM |
| Number of images | 6 | 4 | 174 | 173 | 31 | 9 | 2 | 1 |
| Images with full optic disc | 0% | 0% | 94.8% | 100% | 100% | 100% | 100% | 100% |
| Images with atrophy | 100% | 75% | 98.3% | 52.6% | 100% | 77.8% | 100% | 0% |
| Images with fovea | 100% | 100% | 99.4% | 100% | 96.8% | 88.9% | 100% | 100% |
| Optic disc area | - | - | 1.66% | 1.72% | 3.38% | 2.61%** | 1.69% | 1.15% |
| Atrophy area | 5.93% | 0.41%* | 11.77% | 0.25%**** | 13.97% | 0.70%**** | 42.37% | - |
| Fovea x mean | 768 | 758 | 1236 | 1102**** | 1261 | 1387* | 1748 | 1792 |
| Fovea y mean | 713 | 741 | 1026 | 1081**** | 754 | 715 | 1144 | 1049 |

*Table 3: Quantitative results on the official PALM validation and test set. (*The fovea localization on validation set was obtained prior to the development of the final postprocessing procedure)*

| Task | Metric | **Validation** (400 images) | **Test** (400 images) |
|---|---|---|---|
| Pathological myopia detection | AUC | 0.9934 | 0.9867 |
| Fovea localization | Euclidean distance | 79.42* | 58.27 |
| Optic disc segmentation | Dice (segmentation) | 0.9418 | 0.9303 |
|  | F1 (detection) | 0.9883 | 0.9869 |
| Atrophy segmentation | Dice (segmentation) | 0.8068 | 0.8001 |
|  | F1 (detection) | 0.9336 | 0.9135 |
| Detachment segmentation | Dice (segmentation) | 0.7086 | 0.8073 |
|  | F1 (detection) | 0.6667 | 0.7059 |

*Table 2: Quantitative results of experiments on holdout validation set of 40 images of the training set.*

| Task (metric) | Experiment | Holdout (40 images) |
|---|---|---|
| PM detection (AUC) | vanilla | 1 (loss: 0.1446) |
| | combined in ONH/atrophy segmentation | **1 (loss: 0.0824)** |
| ONH segmentation (Dice) | segmentation, vanilla | **0.9481** |
| | segmentation, combined with atrophy | 0.9462 |
| | segmentation, combined with atrophy, Lovász | 0.9414 |
| Atrophy segmentation (Dice) | segmentation, vanilla | 0.6210 |
| | segmentation, combined with ONH | 0.6810 |
| | segmentation, combined with atrophy, Lovász | **0.6948** |
| Fovea localization (Euclidean distance) | regression with Euclidean distance loss | 229.428 |
| | segmentation, radius 25 pixels | 129.182 |
| | segmentation, radius 75 pixels | 109.770 |
| | segmentation, cutout, combined with ONH | 86.675 |
| | segmentation, cutout, combined with ONH, postprocessing | **61.924** |
| Detachment segmentation (Dice) | segmentation, vanilla | 0.9500 |
| | segmentation, balanced data generator | **0.9998** |